\documentclass[%
aps,
prd,
twocolumn,
superscriptaddress,
showpacs,
floatfix,
amsmath,
amssymb,
longbibliography,
10pt
]{revtex4-1}

\usepackage[utf8]{inputenc}
\usepackage[T1]{fontenc}
\usepackage{txfonts}
\usepackage{tgtermes}
\usepackage[altg]{qtxmath}
\usepackage{braket}
\usepackage{graphicx}
\graphicspath{{figures/}}
\usepackage{color}
\usepackage[english]{babel}
\usepackage{microtype}
\usepackage{lipsum}
\usepackage{url}
\usepackage[breaklinks=true,
pdfauthor={Q.~Z. Lv},
pdftitle={Enhancement of pair creation due to locality in bound-continuum interactions}]{hyperref}

\newcommand*{\llangle}{\langle\!\langle}
\newcommand*{\rrangle}{\rangle\!\rangle}
\def\bbra#1{\mathinner{\llangle{#1}\|}}
\def\kket#1{\mathinner{\|{#1}\rrangle}}

\def\BBraket#1{\left\langle\!\left\langle#1\right\rangle\!\right\rangle}
{\catcode`\|=\active
  \xdef\BBraket{\protect\expandafter\noexpand\csname BBraket \endcsname}
  \expandafter\gdef\csname BBraket \endcsname#1{\begingroup
     \ifx\SavedDoubleVert\relax
       \let\SavedDoubleVert\|\let\|\BraDoubleVert
     \fi
     \mathcode`\|32768\let|\BraVert
     \left\langle\!\left\langle{#1}\right\rangle\!\right\rangle\endgroup}
}

\renewcommand*{\vec}[1]{\boldsymbol{#1}}
\newcommand*{\D}{\mathrm{d}}
\newcommand*{\I}{\mathrm{i}}

\newcommand*{\op}[1]{{\hat#1}}
\newcommand*{\trans}[1]{#1{}^{\mathsf{T}}}

\newcommand*{\acommut}[2]{\mathchoice{\left\{#1,#2\right\}}{\{#1,#2\}}{\{#1,#2\}}{\{#1,#2\}}}
\DeclareMathOperator*{\V}{|}

\begin{document}

\title{Enhancement of pair creation due to locality in bound-continuum interactions}

\affiliation{Beijing National Laboratory for Condensed Matter Physics,
  Institute of Physics, Chinese Academy of Sciences, 100190 Beijing, China}
\affiliation{School of Physical Sciences, University of Chinese Academy
  of Sciences, 100190 Beijing, China}
\affiliation{Max-Planck-Institut f{\"u}r Kernphysik, Saupfercheckweg 1, 69117 Heidelberg, Germany}
\affiliation{Songshan Lake Materials Laboratory, 523808 Dongguan, Guangdong, China}
\affiliation{Key Laboratory for Laser Plasmas (Ministry of Education), School of Physics and
  Astronomy, Shanghai Jiao Tong University, 200240 Shanghai, China}

\author{D.~D. Su}
\affiliation{Beijing National Laboratory for Condensed Matter Physics,
  Institute of Physics, Chinese Academy of Sciences, 100190 Beijing, China}
\affiliation{School of Physical Sciences, University of Chinese Academy
  of Sciences, 100190 Beijing, China}

\author{Y.~T. Li}
\affiliation{Beijing National Laboratory for Condensed Matter Physics,
  Institute of Physics, Chinese Academy of Sciences, 100190 Beijing, China}
\affiliation{School of Physical Sciences, University of Chinese Academy
  of Sciences, 100190 Beijing, China}
\affiliation{Songshan Lake Materials Laboratory, 523808 Dongguan, Guangdong, China}

\author{Q.~Z. Lv}
\email{qingzheng.lyu@mpi-hd.mpg.de}
\affiliation{Max-Planck-Institut f{\"u}r Kernphysik, Saupfercheckweg 1, 69117 Heidelberg, Germany}

\author{J. Zhang}
\affiliation{Key Laboratory for Laser Plasmas (Ministry of Education), School of Physics and
  Astronomy, Shanghai Jiao Tong University, 200240 Shanghai, China}

\date{\today}

\begin{abstract}
  { Electron-positron pair production from vacuum is studied in
    combined background fields, a binding electric potential well and
    a laser field.  The production process is triggered by the
    interactions between the bound states in the potential well and
    the continuum states in the Dirac sea.  By tuning the binding
    potential well, the pair production can be strongly affected by
    the locality of the bound states.  The narrower bound states in
    position space are more efficient for pair production.  This
    is in contrast to what is commonly expected that the wider
    extended bound states have larger region to interact with external
    fields and would thus create more particles.  This surprise can be
    explained as the more localized bound states have a much wider
    extension in the momentum space, which can enhance the
    bound-continuum interactions in the creation process.  This
    enhancement manifests itself in both perturbative and
    non-perturbative production regimes.  }
\end{abstract}


\maketitle

\section{Introduction}
\label{sec:intro}

The vacuum state is the lowest energy states of a quantum
electrodynamics(QED) system in a field-free background.  However,
there exist certain classes of electromagnetic fields in which the
quantum vacuum can become unstable as electron-positron pair
production occurs
\cite{Di-Piazza:Muller:etal:2012:Extremely_high-intensity}.  While
early predictions of this possibility date back to Heisenberg and
Euler \cite{Heisenberg:Euler:1936:Folgerungen_aus}, Sauter
\cite{Sauter:1931:Uber_Verhalten} in the beginning part of last
century, this subject has attracted sustained interest from both
theoreticians and experimentalists in recent years because of the
corresponding experimental studies planned at upcoming high-intensity
laser facilities, such as the Extreme-Light-Infrastructure
\cite{website:eli,dunne:2009:new}, the Exawatt Center for Extreme
Light Studies \cite{website:xcels} or the European X-Ray Free-Electron
Laser \cite{website:hibef,ringwald:2001:paircreation_X-ray}.

The first calculation of the pair production rate in a static
homogeneous electric field based on a non-perturbative approach was
accomplished by Schwinger \cite{Schwinger:1951:On_Gauge} in the early
1950s, according to which a sizeable pair-creation rate requires a
field $E_\mathrm{cr} = m_\mathrm{e}^2 c^3/(e \hbar) = 1.3 \times
10^{18}\,\mathrm{V/m}$, which is still beyond the current technology.
Here $m_\mathrm{e}$, $e$ and $c$ denote the electron mass, the
elementary charge and the speed of light.  $\hbar$ is the reduced
Planck constant.  In order to realize the pair creation below the
critical field strength $E_\mathrm{cr}$, the follow up studies
\cite{Hansen:Ravndal:1981:Kleins_Paradox,Holstein:1998:Kleins_paradox}
have extended Schwinger’s pioneering work to calculate the long-time
pair creation behavior in background fields with more general
configurations.

One of the most important extensions is the so-called dynamically
assisted Schwinger process
\cite{Schutzhold:Gies:etal:2008:Dynamically_Assisted,krajewska2019unitary,orthaber2011momentum,otto2015lifting,linder2015pulse,otto2015dynamical,schneider2016dynamically,torgrimsson2017dynamically,torgrimsson2018sauter,aleksandrov2018dynamically},
where a strong slowly varying field is combined with a weak but highly
oscillating field.  In this circumstance, the underlying mechanism is
a mixture of tunneling process and multi-photon process
\cite{di2009barrier} and thus the pair production probability can be
strongly enhanced.  In addition to this mixture mechanism, several
recent investigations involve also colliding directly two or more
laser pulses to create electron-positron pairs through pure
multi-photon processes
\cite{Ruf:Mocken:etal:2009:Pair_Production,Bulanov:Mur:etal:2010:Multiple_Colliding,Wollert:Bauke:etal:2015:Spin_polarized,jansen2016strong}.
Other studies also investigate the production of electron-positron
pairs in the combination of general electric and magnetic fields
\cite{gavrilov2019pair,su:2012:magnetic,kohlfurst:2018:phase}.  The
creation process in a thermal back ground is also considered recently
using worldline instanton technique, see
\cite{medina2017schwinger,korwar2018finite,brown2018schwinger,wang2019schwinger}
and the references therein.

Nowadays, physicists commonly believe that by choosing the appropriate
field configurations both in space- and time-domains one can amplify
the pair production
\cite{torgrimsson:2019:perturbative,taya:2019:franz,otto2015dynamical,torgrimsson2016doubly,kohlfurst2013optimizing,dong2017optimization,hebenstreit2014optimization,unger2019optimal}.
A well-known procedure is to employ the bound states
\cite{Jiang:Lv:etal:2013:Enhancement_of,Fillion-Gourdeau:Lorin:etal:2013:Resonantly_Enhanced,fillion:2013:enhanced,Tang:Xie:etal:2013:Electron-positron_pair,wang:2016:pumping,Wang:2019:elepos}
in some binding potentials as the bridge between the positive and
negative energy states to enhance pair production.  This can be
realized in laboratory by shooting a laser at a highly charged ion or
nucleus.  However, it remains unknown how the properties of the bound
states affect the pair production process.  For instance, will the
creation rate be increased or decreased due to the localization of the
bound state?  Locality is one of the main characteristics of a bound
state.  Naively speaking, a more extended bound state in position
space will provide a large chance to interact with the external fields
and thus contribute more to the production.  Nevertheless, we will
show in this paper that the more localized bound states actually
enhance the pair creation.

On the other hand, the energy of the bound states plays a major role
in the pair creation processes induced by bound-continuum interaction
\cite{Fillion-Gourdeau:Lorin:etal:2013:Resonantly_Enhanced,fillion:2013:enhanced,Tang:Xie:etal:2013:Electron-positron_pair}.
However, to the best of our knowledge, there is no examination of
whether the required energy conservation being the only criterion for
the pair production to be triggered.  Both issues will be addressed in
this article, which focuses on the pair creation caused by an external
binding potential with or without a laser field.

We study the pair production by employing the computational quantum
field theory (CQFT) approach
\cite{cheng:2010:introductory,xie:2018:electron,Lv:2017:time}.  Two
complementary regimes are considered.  We begin with assuming that the
binding potential well is subcritical and the bound states appear in
the energy gap.  A laser field is then superimposed onto the potential
well and triggers the transition between Dirac sea and bound states.
This situation can be treated perturbatively as the laser field is a
small perturbation.  Secondly, we also investigate pair creation when
the binding potential is supercritical.  Here the quasi-bound states,
caused by the true bound states embedded in the Dirac sea, can
exclusively induce pair production and a laser field is not necessary.
We will demonstrate that a more localized bound state can enhance pair
creation in both cases.  Furthermore, we will show that the energy of
the bound states is not the only condition that determines the pair
production rate.  After the energy conservation law is fulfilled, the
locality of the bound states plays a more important role.

This paper is organized as follows.  In order to render the
presentation self-contained, Section \ref{sec:theo} is devoted to a
concise review of the theoretical framework of the computational
quantum field theory, which allows us to investigate the pair-creation
dynamics with space-time resolutions in arbitrary external force
fields.  In Section \ref{sec:bound_continuum}, we give an intuitive
picture of the two different regimes for the pair creation process.
The enhancement of the pair production caused by the localization of
the bound states is investigated in both perturbative interaction
regime (Sec.~\ref{sec:perturbative}) and non-perturbative regime
(Sec.~\ref{sec:nonperturbative}).  In Section \ref{sec:Summa}, we give
a brief summary and an outlook for further studies.

\section{The theoretical framework of computational quantum field theory}
\label{sec:theo}

In order to describe the dynamics of pair production process, the
relativistic quantum mechanical (Dirac) equation for a single-particle
wave function is not sufficient as its unitary time evolution would
preserve the number of particles in the system.  To describe creation
and annihilation processes we need the time-dependence of the field
operator, which can be obtained from solving the Heisenberg equation
of motion using the quantum field theoretical Hamiltonian.  However,
as we use the strong field approximation where the interfermionic
interaction is neglected and the external fields are treated
classically, it turns out that the Heisenberg equation is equivalent
to the Dirac equation \cite{Geiner:Reinhardt:1996:Field_quantization}
\begin{equation}
  \label{eq:dirac_equ}
  \I\hbar \frac{\partial \op \Psi(\vec r, t)}{\partial t} =
  \op{H}_\mathrm{D}\op\Psi(\vec r, t) \, ,
\end{equation} 
with the Hamiltonian operator
\begin{equation}
  \label{eq:hamiltonian_dirac}
  \op{H}_\mathrm{D} =
  c \vec\alpha\cdot(\vec{\op p} -q \vec A(\vec r, t))  +
  \beta m_\mathrm{e}c^2 + q\phi(\vec r, t)
\end{equation}
Here, we also introduced the momentum operator $\vec{\op p}$, the
charge for an electron $q=-e$, as well as the Dirac matrices
$\vec{\alpha}=\trans{(\alpha_1, \alpha_2, \alpha_3)}$ and $\beta$.
The background fields here are represented by the electromagnetic
scalar potential $\phi(\vec r, t)$ and vector potential $\vec A(\vec
r, t)$.  The field operator $\op \Psi(\vec r, t)$ can be expanded in
terms of two different sets of creation and annihilation operators as
follows:
\begin{equation}
  \label{eq:field_operator}
  \begin{split}
    \op\Psi(\vec r, t) = &  \sum_{\vec p, s} \op b_{\vec p, s}(t) \psi_{\vec p, s}^+(\vec r) +
                      \sum_{\vec p, s} \op d^{\ \dagger}_{\vec p, s}(t) \psi_{\vec p, s}^-(\vec r) \\
                    = & \sum_{\vec p, s} \op b_{\vec p, s} \psi_{\vec p, s}^+(\vec r, t) +
                      \sum_{\vec p, s} \op d^{\ \dagger}_{\vec p, s} \psi_{\vec p, s}^-(\vec r, t) \,.
   \end{split}               
\end{equation}
Here, $\psi_{\vec p, s}^+(\vec r)$ denotes a normalized free-particle
state with positive energy and momentum eigenvalue $\vec p$ and spin
$s$, and correspondingly $\psi_{\vec p, s}^-(\vec r)$ denotes a
free-particle state with negative energy, while the functions
$\psi_{\vec p, s}^+(\vec r, t)$ and $\psi_{\vec p, s}^-(\vec r, t)$
denote the solutions of the time-dependent Dirac equation with
$\psi_{\vec p, s}^+(\vec r)$ and $\psi_{\vec p, s}^-(\vec r)$,
respectively, as initial conditions at time $t=0$.  The fermionic
annihilation and creation operators satisfy the anticommutation
relations
\begin{equation}
  \label{eq:anti_comm}
  \begin{split}
   & \acommut{\op b_{\vec p, s}}{\op b^{\ \dagger}_{\vec p', s'}}=
     \acommut{\op d_{\vec p, s}}{\op d^{\ \dagger}_{\vec p', s'}}=
     \delta_{\vec p, \vec p'}\delta_{s, s'}  \\
   & \acommut{\op b_{\vec p, s}(t)}{\op b^{\ \dagger}_{\vec p', s'}(t)}=
     \acommut{\op d_{\vec p, s}(t)}{\op d^{\ \dagger}_{\vec p', s'}(t)}=
     \delta_{\vec p, \vec p'}\delta_{s, s'}  \,,
  \end{split}
\end{equation}
where $\delta_{i, j}$ denotes a Kronecker delta.  All other
anticommutators are zero.  We can, now, equate the time dependent
creation and annihilation operators with the time independent ones
through the generalized Bogoliubov transformation, for example,
\begin{equation}
  \op b_{\vec p, s}(t) =
  \sum_{\vec p', s'}
  G_{\vec p, s; \vec p', s'}(\sideset{^+}{_{+}}\V)\op b_{\vec p', s'} +
  G_{\vec p, s; \vec p', s'}(\sideset{^+}{_{-}}\V)\op d^{\ \dagger}_{\vec p', s'}
\end{equation}
and
\begin{equation}
  \op d^{\ \dagger}_{\vec p, s} (t) =
  \sum_{\vec p', s'}
  G_{\vec p, s; \vec p', s'}(\sideset{^-}{_{+}}\V)\op b_{\vec p', s'} +
  G_{\vec p, s; \vec p', s'}(\sideset{^-}{_{-}}\V)\op d^{\ \dagger}_{\vec p', s'}
\end{equation}
with the transition amplitudes
\begin{equation}
  G_{\vec p, s; \vec p', s'}(\sideset{^\nu}{_{\nu'}}\V) =
  \braket{\psi_{\vec p, s}^\nu(\vec r)|\psi_{\vec p', s'}^{\nu'}(\vec r, t)} \,.
\end{equation}

Stripping the antiparticle part from the quantum field
operator~\eqref{eq:field_operator}, the electronic portion of the
field operator associated with positive energy can then be defined as
\begin{equation}
  \label{eq:field_operator_particle}
  \op\Psi_+(\vec r, t) =
  \sum_{\vec p, s} \op b_{\vec p, s}(t)\psi_{\vec p, s}^+(\vec r) \,.
\end{equation}
With this definition operators representing various physical
quantities, can be calculated, e.\,g., the average spatial density of
the created electrons
\begin{equation}
  \label{eq:rho_x}
  \begin{split}
    \varrho(\vec r, t) = &
    \bbra{vac} \op\Psi^{\dagger}_+(\vec r, t)\op\Psi_+(\vec r, t) \kket{vac} \\
    = & \sum_{\substack{\vec p, s \\ \vec p', s'}} S_{\vec p, s;\vec p', s'}(t)
    \psi_{\vec p, s}^{+\ \dagger}(\vec r)\psi_{\vec p', s'}^+(\vec r) \,,
  \end{split}
\end{equation}
and the momentum distribution 
\begin{equation}
  \label{eq:rho_p}
  \chi^-(\vec p, t) =
  \bbra{vac}\sum_{s} \op b^{\ \dagger}_{\vec p, s}(t)\op b_{\vec p, s}(t) \kket{vac}=
  \sum_{\substack{s}} S_{\vec p, s;\vec p, s}(t) \,.
\end{equation}
Here we have introduced the Hermitian matrix
\begin{equation}
  S_{\vec p, s; \vec p', s'}(t) = \sum_{\vec p'', s''}
  G_{\vec p, s; \vec p'', s''}^*(\sideset{^+}{_{-}}\V)
  G_{\vec p', s'; \vec p'', s''}(\sideset{^+}{_{-}}\V)
\end{equation}Then the average number of the created particles can be calculated as
\begin{equation}
  \label{eq:N_t}
  N(t)=\int \D^3 r \ \varrho(\vec r, t) = \int \D^3 p \ \chi(\vec p, t)= \sum_{\vec p, s} S_{\vec p, s; \vec p, s}(t) \,.
\end{equation}  
While $\psi_{\vec p', s'}^{\nu'}(\vec r, t)$ can be obtained by
evolving the Dirac equation numerically with the split-operator
technique
\cite{bandrauk:1993:exponential,braun:1999:numerical,bauke:2011:accelerating},
the matrices $S_{\vec p, s; \vec p', s'}(t)$ are calculable at all
times, as are the spatial density $\varrho(\vec r, t)$, the momentum
spectrum $\chi^-(\vec p, t)$ and the average particle number $N(t)$.

The numerical solution of the corresponding physical quantities on a
space-time grid provides us deeper insight when studying the dynamics
of pair production processes than the standard S-matrix approach,
which can only represent the system’s asymptotic behavior.

\section{Bound-continuum interactions}
\label{sec:bound_continuum}

Before we describe the results, let us first review the physical
picture of two different regimes for the bound-continuum interactions
in the pair production process.  Our goal is to study how the
properties of the bound states in a binding potential play a role in
the pair production process.  For numerical feasibility, we choose a
localized scalar potential well of the form
\begin{equation}
  \label{eq:well_pot}
  q\phi(x, t) = -V_0[S(x+D/2)-S(x-D/2)]f(t)
\end{equation}
instead of the long range Coulomb field.  Here the parameter $D$ is
related to the spatial width of the well, which is formed by two
smooth unit-step functions $S(x) = \frac{1}{2}\left(1 + \tanh \frac x
W \right) $, where $W$ is the extent of the associated localized
electric fields \cite{Sauter:1931:Uber_Verhalten}.  The time dependent
function $f(t)$ is used to imitate the turn-on and turn-off processes of
the external field in experiments.  In our calculation, we have
\begin{equation}
  \label{eq:turn_on_and_off}
  f(t)=
  \begin{cases}
    \sin^2\frac{\pi(t-\Delta T)}{2\Delta T} &
    \text{for $-\Delta T\le t\le 0$}\,, \\
    1 & \text{for $0 \le t \le T$}\,, \\
    \cos^2\frac{\pi (t-T)}{2\Delta T} &
    \text{for $T \le t \le T+\Delta T$} \,,
  \end{cases}
\end{equation}
where $T$ denotes the duration of the flat plateau and $\Delta T$ the
duration for turn-on and turn-off.  The field configuration at the
plateau phase can support several electronic bound states.  These
bound states act like a bridge between negative and positive energy
states in the Dirac sea picture to induce transition between them and
create electron-positron pairs from vacuum.

\begin{figure}
  \centering
  \includegraphics[scale=0.275]{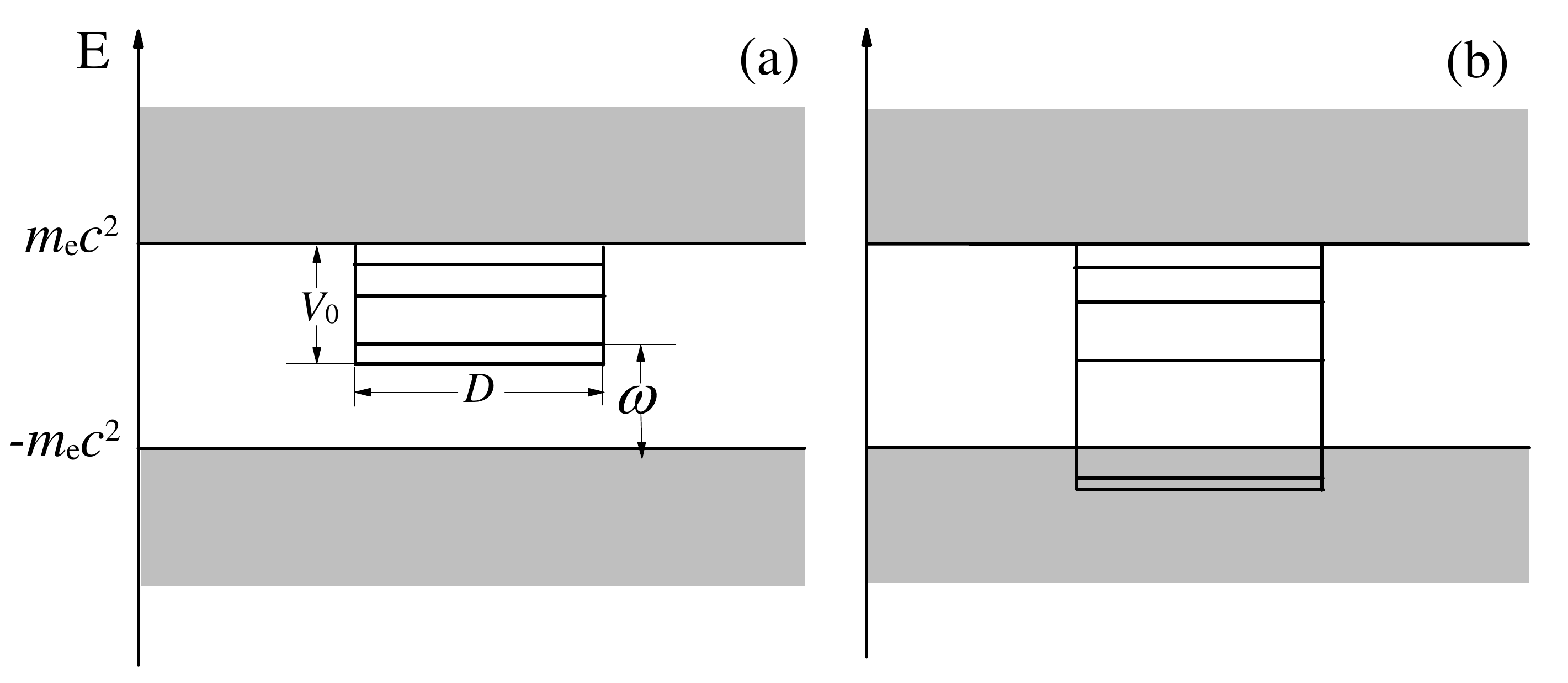}
  \caption{Sketch of the bound-continuum interaction in a binding
    potential well for the pair production processes.  Panel (a) shows
    the perturbative regime in which the bound states appear in the
    energy gap.  Pair production cannot be induced solely by the
    binding potential well and the laser field is necessary.  In the
    non-perturbative regime, panel (b), the bound states overlap with
    the negative-energy continuum.  Particles can be created by
    tunneling of the initially occupied negative-energy states into
    the quasibound states.  Here the laser field is unnecessary and
    the tunneling dynamics cannot be treated perturbatively.}
  \label{fig:pertur_nonpertur}
\end{figure}

With different choices of the potential height $V_0$, it is well-known
that there exist two separate parameter regimes, which have completely
different mechanisms for pair creation.  As in
Fig.~\ref{fig:pertur_nonpertur}, the left panel shows that when
$V_0<2m_\mathrm{e} c^2$, all the bound states are present in the
energy gap and thus no particles can be created alone by this binding
potential.  However, if now a laser field with frequency $\omega$ is
superimposed onto the binding potential well, the pair creation can
then be triggered by the combined fields provided that the energy
conservation law is fulfilled.  Since the intensity of the laser field
needed here is rather weak compared to Schwinger's critical intensity,
it can be viewed as a small perturbation.  This is the regime where
perturbative (multi-photon) mechanism dominates the creation
\cite{Jiang:Lv:etal:2013:Enhancement_of,Tang:Xie:etal:2013:Electron-positron_pair}.

On the other hand, as shown in Panel (b) of
Fig.~\ref{fig:pertur_nonpertur}, the increase of $V_0$ will overlap
the lower bound states with the negative-energy continuum.  The
resulting degeneracy between the quasibound states and the
negative-energy continuum leads to the instantaneous pair creation,
like in the case of the Coulomb field in ion collision experiments.
The production mechanism in this regime is non-perturbative since the
particles are created through tunneling dynamics.  Several interesting
phenomena appear in this regime, like the non-competing mechanism
between different channels
\cite{Lv:Liu:etal:2013:Noncompeting_Channel,Lv:Liu:etal:2014:Degeneracies_of}
when there are more than one quasibound state for the creation and
like that the system will instantaneously evolve into a multi-pair
field-state at the end \cite{Lv:2017:time}.

\section{Enhancement of pair production in the perturbative regime}
\label{sec:perturbative}

In this section, we will study the pair production process in the
perturbative regime.  As mentioned in Sec.~\ref{sec:bound_continuum},
in order to trigger pair creation, we have to superimpose a laser
field onto the subcritical binding potential.  Here we choose a
laser field represented by the vector potential $\vec A(\vec r, t) =
(0, A_0 f(t) \sin \omega (t - x/c), 0)$, where $A_0$ is the amplitude
of the potential and $\omega$ is the frequency.  The time-dependent
envelope $f(t)$ is the same as in Eq.~\eqref{eq:turn_on_and_off} to
characterize the turn-on and turn-off of the field.  

It is well-known that the criterion for pair production is the energy
provided by the external field should be at least equal or larger than
the rest energy of the created particles.  In our case, it means that
the binding energy of the bound state and the energy of the absorbed
laser photons should together be larger than $2m_\mathrm{e}c^2$.  To
study the effect of the locality of the bound states in the production
process, we have chosen the parameters as $V_0=1.726 m_\mathrm{e}c^2$
and $D=3.200 \lambda_c$ as well as $V_0=1.900 m_\mathrm{e}c^2$ and
$D=2.443 \lambda_c$, so that the energies of the ground states in both
potential wells are the same $E_\mathrm{g}=-0.4m_\mathrm{e}c^2$.  Here
$\lambda_c$ denotes the Compton wave length for the electron.  The
frequency of the laser is $\omega=0.45 m_\mathrm{e}c^2$, which means
that the energy of the laser photon is $0.45 m_\mathrm{e}c^2$, and the
amplitudes $A_0$ is chosen such that the electric field is $E_0= A_0
\omega/c=0.3 E_\mathrm{cr}$.  Since the electric field for these
parameters is much smaller than the critical field, the
non-perturbatively tunneling pair creation is suppressed and electrons
can only be transmitted perturbatively into the ground state by
absorbing two or more photons from the laser.

\begin{figure}
  \centering
  \includegraphics[scale=0.10]{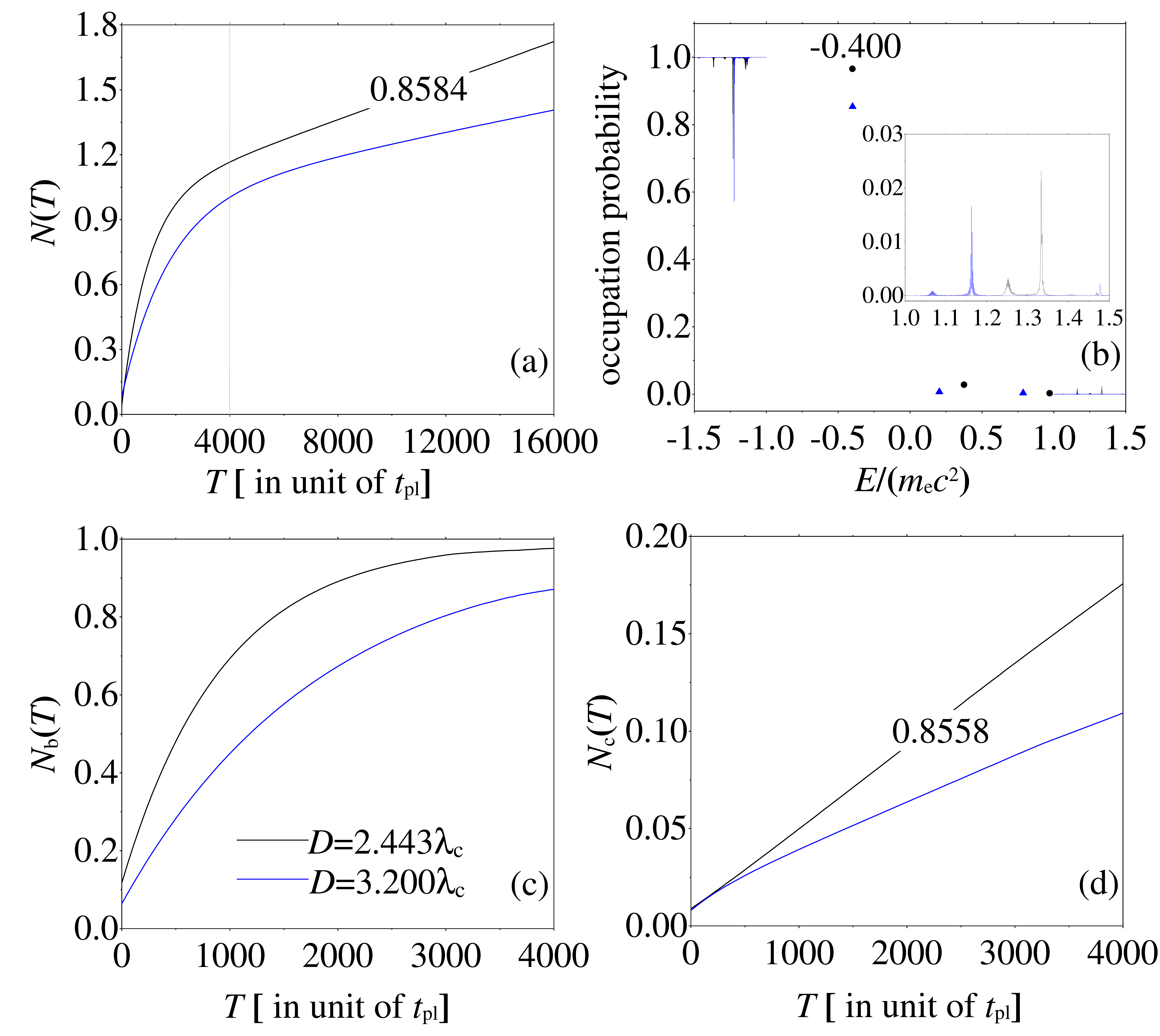}
  \caption{Figures for the pair production process in two different
    binding potential wells.  Panel (a) is the average particle number
    $N(T)$ as a function of interaction time $T$.  Here $t_\mathrm{pl}
    = \lambda_c/c$ denotes the typical time scale for the pair
    production process.  Panel (b) is the probability of the
    occupation of the instantaneous state at $T=480 \pi / \omega =
    3349 t_\mathrm{pl}$.  Panel (c) and (d) are the average particle
    number in the instantaneous ground state and positive continuum,
    respectively, as a function of time $T$.  The color is coded as
    blue for the potential with $V_0=1.726 m_\mathrm{e}c^2$ and
    $D=3.200 \lambda_c$ and black for $V_0=1.900 m_\mathrm{e}c^2$ and
    $D=2.443 \lambda_c$.  The other parameters are $W=0.3
    \lambda_c$for the potential wells and $\omega=0.45
    m_\mathrm{e}c^2$ and $E_0=A_0 \omega/c=0.3 E_\mathrm{cr}$ for the
    laser field.}
  \label{fig:NT_OB_NbT_NcT}
\end{figure}

Since the external field used here cannot couple different spin
states, we have considered only a certain spin direction (along $+z$
direction) in the simulations.  Panel (a) in
Fig.~\ref{fig:NT_OB_NbT_NcT} shows the particle number $N(T)$ in the
two potential wells as a function of the interaction time $T$.  As
most of the created electrons occupy the ground state, the average
particle number should, at the end of the interaction, reach unity
because of the Pauli exclusive principle.  The figure, however, shows
that the particle number finally exceeds unity and tends to increase
linearly.  This long time linear increase is a bit surprising since
the bound-continuum interactions cannot induce a permanent pair
creation.

To understand this linear creation and also prove our assumption that
the created particles should mainly occupy the ground state in the
potential well, we have, in Panel (b) of Fig.~\ref{fig:NT_OB_NbT_NcT},
displayed the occupation probability of the instantaneous states after
the creation.  Here the instantaneous states denote the eigenstates of
the Hamiltonian of Eq.~\eqref{eq:hamiltonian_dirac} with only the
binding potential as the background field.  The details of the method
can be found in Ref.~\cite{liu:2014:population}.  The almost $100 \%$
occupation of the negative-energy continuum is consistent with that
the vacuum state means all the negative-energy states being occupied.
This is because the potential well here is subcritical and the
structure of the vacuum state with or without the background potential
well is similar.

Two aspects of the graph deserve further attention.  First of all,
despite most of the negative-energy states being fully occupied, there
is a large peak in the negative continuum showing that these
particular states are much less occupied.  The position of this peak
is around $-1.23 m_\mathrm{e}c^2$ for both cases.  These depopulated
states are caused by the two-photon transition of the Dirac sea states
into the ground state.  This peak also consists with the energy of the
created positrons shown below.

Secondly, the most occupied bound state in the energy gap is the
ground states in both cases with energy $E_\mathrm{g} = -0.4
m_\mathrm{e}c^2$ and all the occupation of the other bound states is
negligible.  This proves our conjecture that the production, in the
earlier time domain, is dominated by the created electrons occupied
the ground state in the potential well.  What is more interesting is
that there are also peaks in the positive continuum.  These small
peaks, which will increase with time, may be the reason of the linear
increase in the particle number $N(T)$ (Panel (a)) for long
interaction time.

In order to test this hypothesis, we have in Panel (c) and (d) of
Fig.~\ref{fig:NT_OB_NbT_NcT} shown the average particle number
$N_b(T)$ occupied the instantaneous ground state $\psi^-(\vec r)$ and
the average particle number $N_c(T)$ in the positive continuum,
respectively.  From the graphs we can see that the population of the
bound states tends to $1$ at the end while the population of the
positive continuum is linearly growing in time.  The sum of $N_b(T)$
and $N_c(T)$ in Panel (c) and (d) approximately equals to the total
average particle number in Panel (a).  More important, the linearly
growing rates of $N_c(T)$ in Panel (d) match the slop of the total
$N(T)$ curves in long interaction time $T$.  For instance, the slop of
the black curve in Panel (d) is about $0.8558$, which differs less
than $1\%$ with the slop ($0.8584$) of the black curve in Panel (a)
for long interaction time.

\begin{figure}
  \centering
  \includegraphics[scale=0.230]{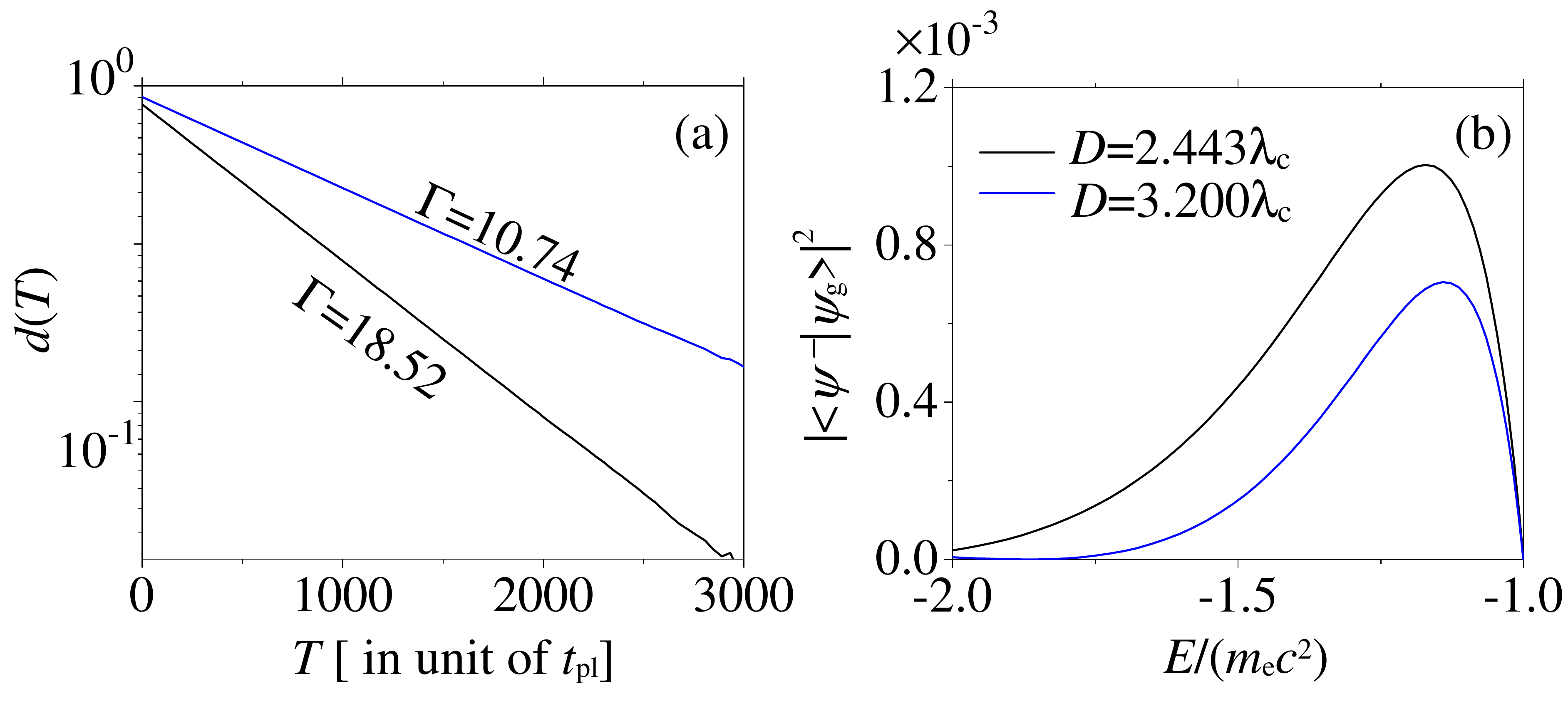}
  \caption{Decay probability $d(T)$ as a function of the interaction
    time $T$ is shown in Panel (a) on a logarithmic scale.  The decay
    rate for the blue line is $\Gamma = 10.74$ while for the black one
    is $\Gamma = 18.52$.  The projection of the ground state in the
    binding potential well onto the field-free negative energy states
    is shown in Panel (b).  Here the ground state
    $\psi_\mathrm{g}(\vec r)$ is the lowest energy eigenstate in the
    energy gap of the Hamiltonian of Eq.~\eqref{eq:hamiltonian_dirac}
    with only the binding potential as the background field and the
    field-free negative continuum is $\psi^-(\vec r)$.  The parameters
    are the same as in Fig.~\ref{fig:NT_OB_NbT_NcT}. }
  \label{fig:decayrate_overlapneg}
\end{figure}

In the early stage of the production, the process is dominated by the
creation of particles in the ground state and from
Fig.~\ref{fig:NT_OB_NbT_NcT}(c) it is obvious that the particle number
in the ground state reaches unity at different speeds.  The black
curve, which is for potential well with $V_0=1.900 m_\mathrm{e}c^2$
and $D=2.443 \lambda_c$, has a larger speed than the blue one.

To provide a more quantitative analysis, we define $d(T)$ as
\begin{equation}
  \label{eq:decay_rate}
  d(T) = | 1 - N_b(T) |\,.
\end{equation}
It characterizes how fast the initial vacuum state decays into
electron-positron pairs in the external fields through the ground
state.  Fig.~\ref{fig:decayrate_overlapneg}(a) shows the quantity
$d(T)$ for the two different cases on a logarithmic scale.  The two
straight lines indicate that the decay process is exponential, namely
$d(T) \propto \exp(-\Gamma T)$ with the exponential parameter $\Gamma$
called the decay rate.

The vacuum decays much faster ($\Gamma=18.52$) in the more localized
system with $D=2.443 \lambda_c$.  This is rather unexpected as it is
commonly believed that the wider the state in position space, the
larger the interaction region and thus the greater the possibility.
In order to understand this counter-intuitive phenomenon, we have to
analyze the properties of the bound states in the two potential wells.

The bound states acting like a bridge in the energy gap can help to
induce pair production in this perturbative regime.  Even it is not
directly related to the creation rate, the nonzero overlap probability
between the bound states and the field-free negative-energy states in
the Dirac sea (shown in Fig.~\ref{fig:decayrate_overlapneg}(b)) could
still help us understand the process more intuitively.  When the
overlap is large, the originally occupied field free Dirac sea states
would be easier to transmit into the bound state in the presence of
the laser field, see also the calculations using time-dependent
perturbation theory in \cite{Jiang:Lv:etal:2013:Enhancement_of}.  From
the figure, it is clear that the more localized bound state (in the
potential well of $V_0=1.900 m_\mathrm{e}c^2$ and $D=2.443 \lambda_c$)
has a larger overlap with the negative-energy continuum.  This is
consistent with the larger decay rate of $\Gamma=18.52$ for the black
line in Fig.~\ref{fig:decayrate_overlapneg}(a).

Please note also that the long time linear creation rate shown in
Fig.~\ref{fig:NT_OB_NbT_NcT}(a) has similar behavior as the decay rate
$\Gamma$ of the vacuum through the bound states for short interaction
time.  This means that the more localized system with $D=2.443
\lambda_c$ creates particles faster in all interaction time region.

All the simulations above are for the case that only ground state is
excited in the interaction as the laser frequency is optimized for
this transition only.  If we want to excite other bound states, we
have to either vary laser frequency or tune the potential well to make
the energy difference between ground state and other bound state
resonant with this laser frequency.  In this circumstance, the other
bound state can be excited even before the particle in the ground
state saturates.

\begin{figure}
  \centering
  \includegraphics[scale=0.227]{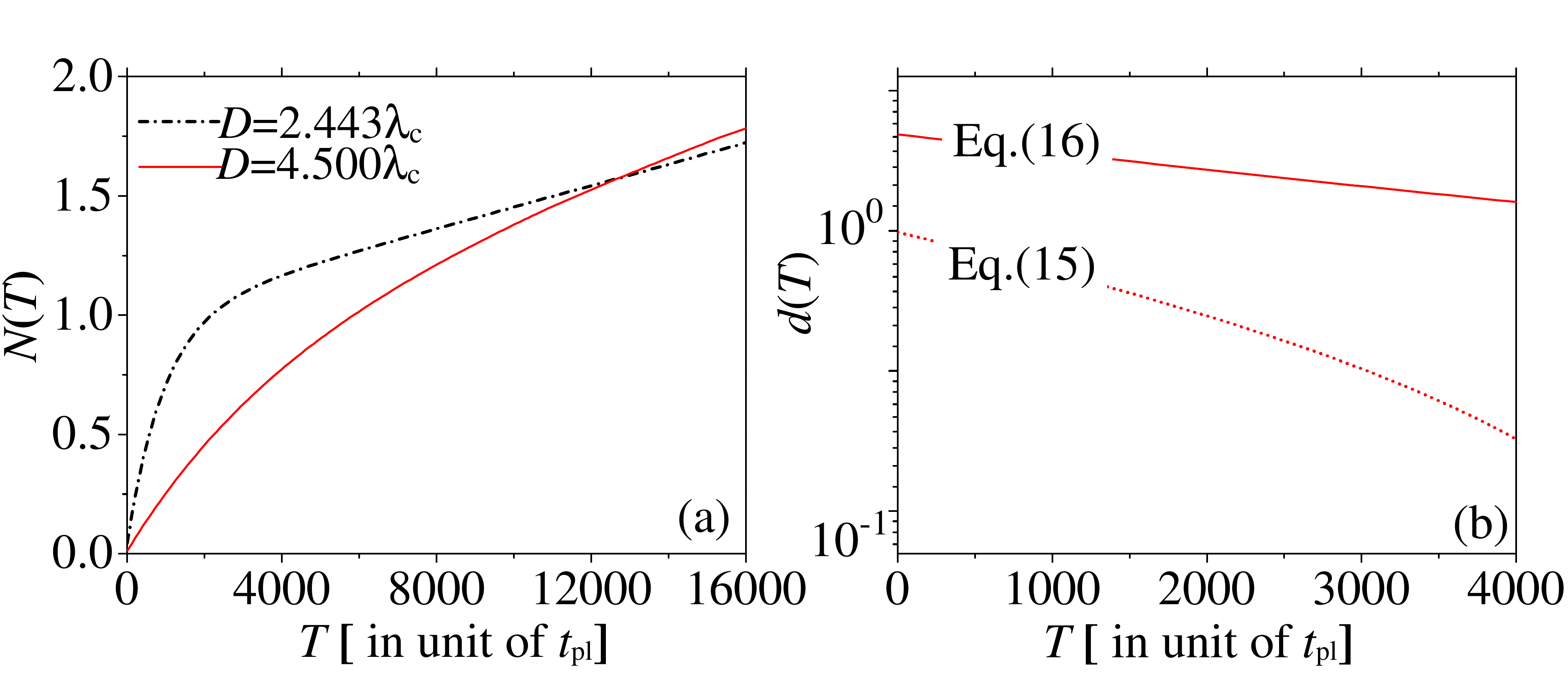}
  \caption{Particle number $N(T)$ and Decay probability $d(T)$ as a
    function of the interaction time $T$ for the case of two bound
    states involved in the pair creation is shown in Panel (a) and
    (b), respectively.  For comparison, the dash-dot line in Panel (a)
    is the replot of the black curve in
    Fig.~\ref{fig:NT_OB_NbT_NcT}(a).  In Panel (b), the decay
    probability with both definitions in Eq.~\eqref{eq:decay_rate2}
    (solid) and Eq.~\eqref{eq:decay_rate} (dotted) is shown.  The
    parameters for the potential well are $V_0=1.584 m_\mathrm{e}c^2$
    and $D=4.500 \lambda_c$ and the laser field are the same as
    before.}
  \label{fig:two_state_excitation}
\end{figure}

In Fig.~\ref{fig:two_state_excitation}, we show the creation for such
a case by change the potential well to ensure the energy difference
between ground state and first excited state $\triangle E \approx
\omega$.  From Fig.~\ref{fig:two_state_excitation}(a), we see that
there is no linear increase before the particle number approaching
$2$.  This is quite different compared with the $N(T)$ curve in
Fig.~\ref{fig:NT_OB_NbT_NcT}(a).  For comparison we have replotted
the black curve of Fig.~\ref{fig:NT_OB_NbT_NcT}(a) as dash-dot line
here.  Since two bound states are excited here, the particle number
will saturate at $2$ and if we want to define a decay rate as in
Fig.~\ref{fig:decayrate_overlapneg}(a), the definition of $d(T)$ has to
be modified like
\begin{equation}
  \label{eq:decay_rate2}
  d(T) = | 2 - N(T) |\,.
\end{equation}
The definition in Eq.~\eqref{eq:decay_rate} is not suitable as shown
in Panel(b) of Fig.~\ref{fig:two_state_excitation}, where the dotted
line is not a linear behavior in the logarithmic scale plot.  However,
the modified $d(T)$, as expected, recovers the decay processes with
the rate $\Gamma=2.259$.

In this situation, the influence of the bound state's extension on
pair creation is more complicated since two bound states are excited.
To concentrate on the fundamental mechanism and keep the discussion
simple, we will in the following discussions still focus on the
creation only involving the ground state.

\begin{figure}
  \centering
  \includegraphics[scale=0.30]{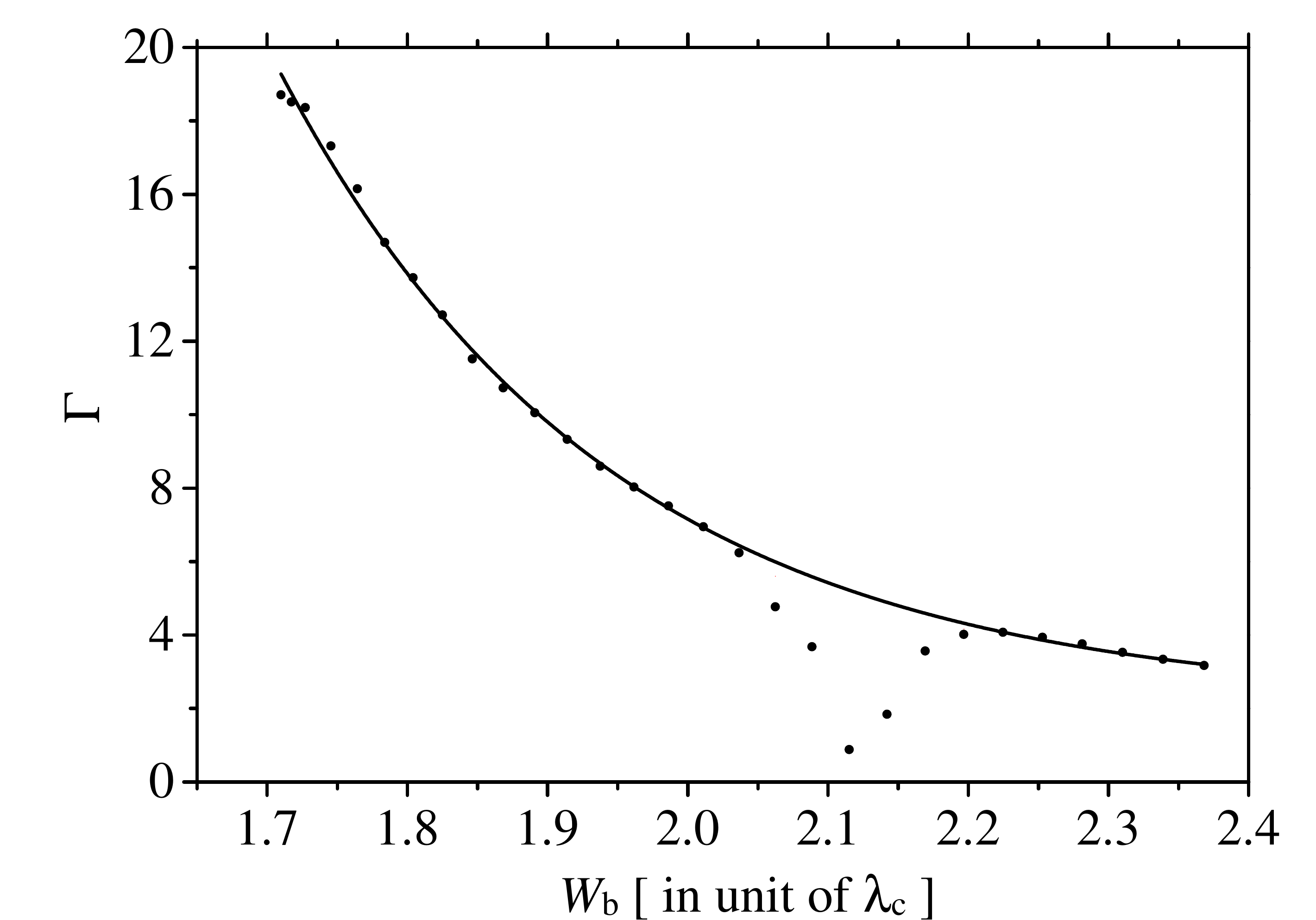}
  \caption{The decay rate $\Gamma$ as a function of the ground state
    width $W_b$.  For comparison, we have chosen the parameters such
    that the energy of the ground states is unchanged $E_\mathrm{g}=
    -0.4m_\mathrm{e}c^2$ for different width of the states.  The
    laser field is the same as in Fig.~\ref{fig:NT_OB_NbT_NcT}.}
  \label{fig:rate_d}
\end{figure}

Our previous results indicate that the pair creation rate decays with
the extension of the bound state.  This is also illustrated in
Fig.~\ref{fig:rate_d}, where the rate $\Gamma$ is shown as a function
of the width $W_\mathrm{b}$ of the ground state.  Here the width of the
ground state is defined as $W_\mathrm{b}= 2 \sqrt{\langle x^2- \langle
  x \rangle^2 \rangle}$ with $\langle x
\rangle=\bra{\psi_\mathrm{b}(\vec r)} \hat{x}
\ket{\psi_\mathrm{b}(\vec r)}$.

The decay rate shown in Fig.~\ref{fig:rate_d} is exponentially
decaying with increasing width of the ground state, $\Gamma \propto
\exp(-C W_\mathrm{b})$, with the constant $C$ depending on the
parameters of the potential well.  There are several points in the
region of $2.062\lambda_c < W_\mathrm{b} < 2.197 \lambda_c$ in the
figure that are not close to the normal decay trend.  The reason is
that the laser field in these cases happens to be able to cause
resonance transitions between the bound states in the energy gap as
shown in Fig.~\ref{fig:two_state_excitation}.  Because of these
resonance transitions, the population in the ground state will
oscillate in time and the decay rate through this state is not as well
defined as for other parameters.

\begin{figure}
  \centering
  \includegraphics[scale=0.30]{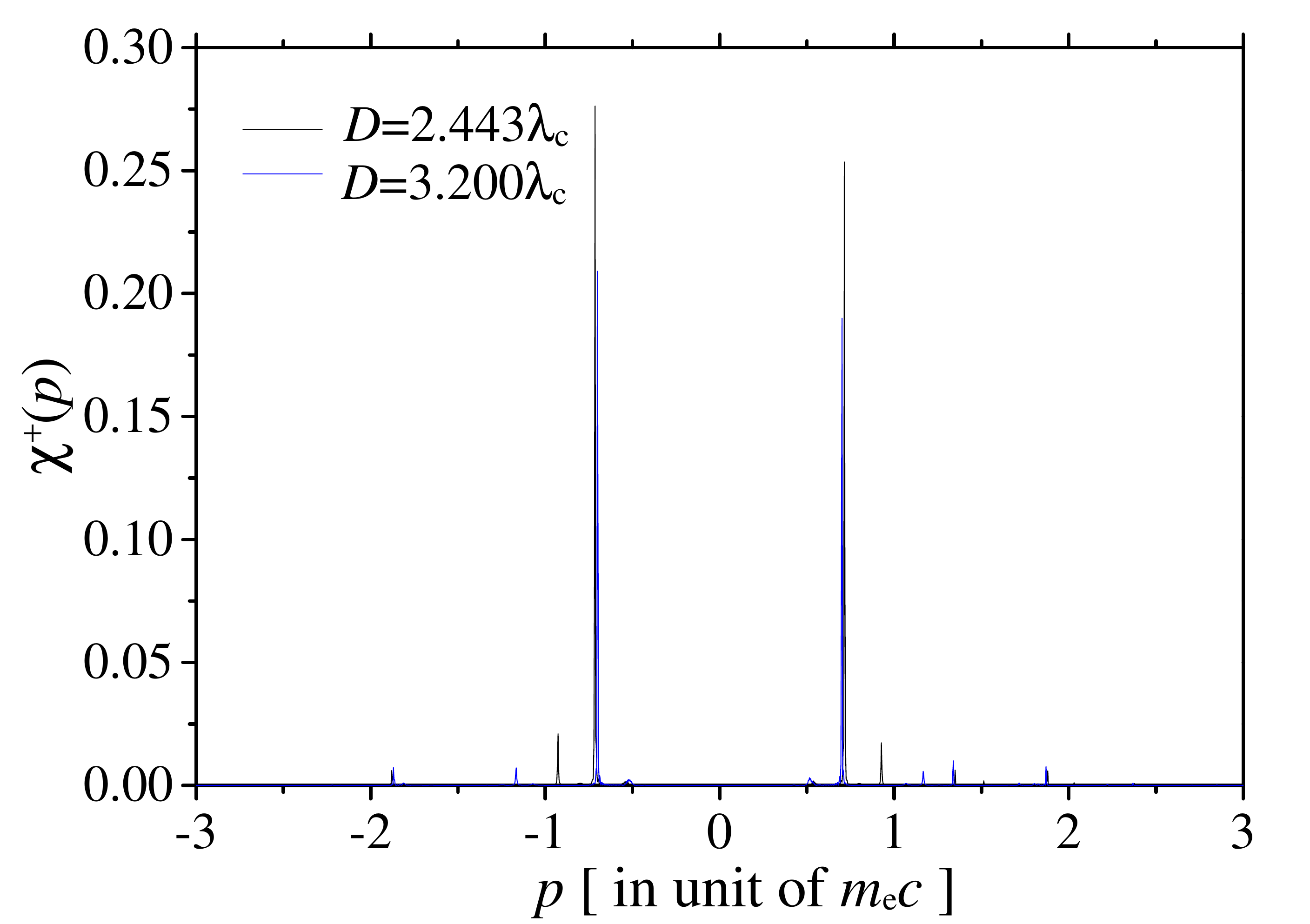}
  \caption{The momentum density of the created positrons at $T=400 \pi
    / \omega = 2791 t_\mathrm{pl}$.  Because of the field configuration
    we use, the only non-conserved momentum during the evolution is
    $p=p_\mathrm{x}$.  For simplicity, we choose here
    $p_\mathrm{y}=p_\mathrm{z}=0$.  The color code and the other
    parameters are the same as in Fig.~\ref{fig:NT_OB_NbT_NcT}.}
  \label{fig:momentum_positron}
\end{figure}

To complete our understanding of the decay process of the vacuum into
electron-positron pairs through the bound-continuum interactions in
perturbative regime, we also investigate the properties of the created
positrons in momentum space.  Unlike the electrons being captured in
the binding potential, the created positrons are free and the momentum
is sharply distributed.  The distribution of the positron in momentum
space $\chi^+(p)$ can be calculated using Eq.~\eqref{eq:rho_p} by
replacing the creation and annihilation operators for electrons to the
operators for positrons.  From Fig.~\ref{fig:momentum_positron}, we
can see that the two main peaks are around $p=\pm 0.71 m_\mathrm{e}c$.
These peaks, if we transfer to energy domain, corresponds to energy of
$1.225 m_\mathrm{e}c^2$, which related to the depopulated states in
the negative-energy continuum in Fig.~\ref{fig:NT_OB_NbT_NcT}(b)
around $E=-1.23 m_\mathrm{e}c^2$.  The small peaks reflect the
acceleration of the positrons in the laser field after the creation.
Since the laser propagates along a certain direction, the momentum
distribution of the positron is not symmetric.

\section{Enhancement of pair production in the non-perturbative regime}
\label{sec:nonperturbative}

In the previous sections, we studied the pair creation in the
perturbative regime.  The results show that the creation can be
enhanced with utilization of a more localized bound state in the
interactions.  In order to complete the picture, we also studied the
production process in a non-perturbative regime in this section.
Unlike in the perturbative regime, we know from
Sec.~\ref{sec:bound_continuum} that the non-perturbative creation is
caused by the diving of the bound states into the Dirac sea as shown
in Fig.~\ref{fig:pertur_nonpertur}(b).

It is known that a quasibound state, a bound state embedded in the
negative continuum, can trigger the decay of the vacuum and thus
produce particle pairs.  Since the quasibound state is not spatially
localized in the continuum, it is not clear if the locality of the
true bound state before diving into the continuum still affects the
pair production.  As the creation is caused by the tunneling of the
Dirac sea states into the initially unoccupied quasibound state, the
perturbative laser field is not necessary here.

\begin{figure}
  \centering
  \includegraphics[scale=0.23]{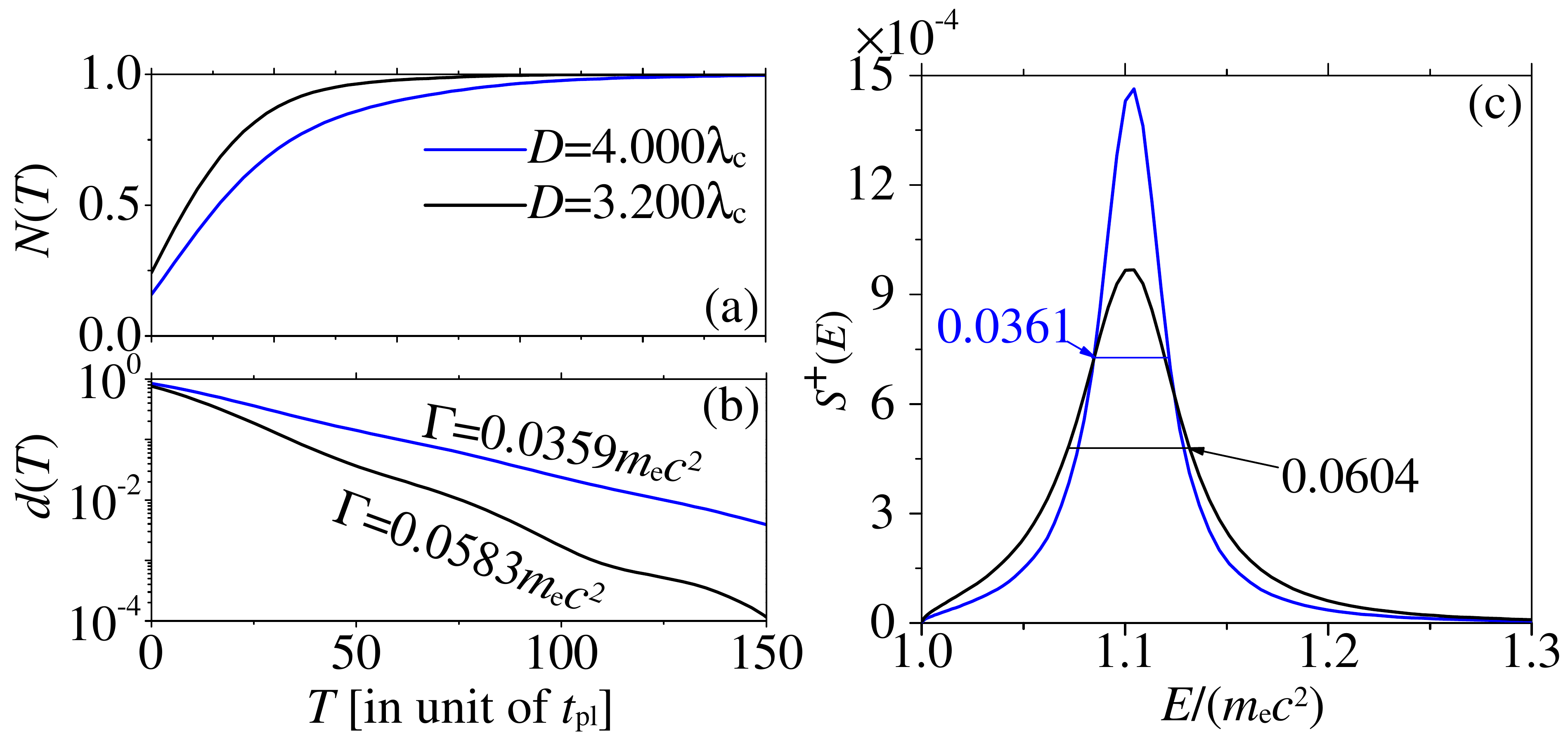}
  \caption{The average particle number $N(T)$ and the decay
    probability $d(T)$ as a function of the interaction time $T$ is
    plotted, respectively, in Panel (a) and (b).  The decay
    probability $d(T)$ is shown in logarithmic scale.  The energy
    spectrum $S^+(E)$ of the created positron for $T=282 t_\mathrm{pl}
    $ is shown in Panel (c).  The parameters for the blue curves are
    $V_0=2.383 m_\mathrm{e}c^2$ and $D=4.000 \lambda_c$ and for the
    black curves are $V_0=2.522 m_\mathrm{e}c^2$ and $D=3.200
    \lambda_c$.  The quasibound states in both potential wells have
    the energy $E_\mathrm{qb}=-1.1 m_\mathrm{e}c^2$.  The choice of
    the parameters insures that the bound states in these two
    potential well have different spatial widths before diving into
    the negative continuum.}
  \label{fig:NT_DT_SE_nonpert}
\end{figure}

In Fig.~\ref{fig:NT_DT_SE_nonpert}(a), we show the average particle
number $N(T)$ for two supercritical potential wells with $V_0=2.383
m_\mathrm{e}c^2$ and $V_0=2.522 m_\mathrm{e}c^2$, respectively.  For
these two cases, the quasibound states are both located at
$E_\mathrm{qb}=-1.1m_\mathrm{e}c^2$.  The graph shows that $N(T)$
tends to one for $T \to \infty$ in contrast to the perturbative case
in Fig.~\ref{fig:NT_OB_NbT_NcT}(a).  This is because that the
electron-positron pairs can only be created through the quasibound
state here.  With one quasibound state, the particle number can only
tend to one eventually.  It is obvious that the particle number tends
to $1$ with different speeds.  To be more quantitative, we also plot
$d(T)$ as defined in Eq.~\eqref{eq:decay_rate} in
Fig.~\ref{fig:NT_DT_SE_nonpert}(b).  The two curves in Panel (b) of
Fig.~\ref{fig:NT_DT_SE_nonpert} indicate that the initial vacuum state
also exponentially decays into electron-positron pairs through the
quasibound state and the potential with $D=3.200 \lambda_c$ triggers
the faster decay.  This is consistent with what happens in the
perturbative regime, for example like in
Fig.~\ref{fig:decayrate_overlapneg}(a).

We know from the previous section that the reason for the
locality-enhancement is that the more localized bound states have more
overlap with the negative continuum.  For the sake of verifying this
explanation in the non-perturbative regime, the energy spectrum
$S^+(E)$ of the created positrons is displayed in
Fig.~\ref{fig:NT_DT_SE_nonpert}(c), which reflects the overlap between
the quasibound state and the Dirac sea states.  $S^+(E)$ is calculated
by transferring the momentum distribution $\chi^+(p)$ to the energy
domain. The two spectra have the similar location for the maximum
value, which corresponds to the energy of the quasibound states.
However, the spectrum for the case of $D=3.200 \lambda_c$ is much
wider than that for $D=4.000 \lambda_c$.  This means that the
quasibound state in the narrower potential well, even it is not
spatially localized, has a larger overlap with the negative continuum.
On the other hand, the full width at half maximum of the two spectra
are consistent with the decay rate in
Fig.~\ref{fig:NT_DT_SE_nonpert}(b).

It is also worth pointing out that the enhancement in this
non-perturbative creation regime might be seen in connection with the
well-known non-Markovian feature of the pair production process
\cite{schmidt:1998:quantum,schmidt:1999:non}, as the quasibound state
inherits some properties from its original bound state.  Because the
ground state in the potential well with $D=3.200 \lambda_c$ is more
localized in the energy gap, its narrow distribution in position space
still amplifies the creation process even after it dives into the
negative Dirac sea and becomes the unlocalized quasibound state.

\section{Summary and outlook}
\label{sec:Summa}

The purpose of this work is to study the influence of the locality of
a bound state in the pair production process.  The feasibility of this
work is the CQFT method, which can give us the full space-time
resolution of the pair production process in any general external
field.  By analyzing the average particle number, we can clearly see
the enhancement of the pair creation process caused by a more
localized bound state. Even with the same binding energy, the vacuum
will decay faster through the bound state with narrower distribution
in position space.  This also means that energy threshold is not the
only criterion for pair creation as some other properties of the bound
states can play a role in the bound-continuum interaction induced pair
production.

The enhancement manifest itself in both perturbative and
non-perturbative regimes, which intrinsically have completely
different mechanisms for triggering pair creation.  In the
perturbative regime, the electron-positron pairs are created by
multi-photon excitation as seen from the momentum spectrum of the
created positrons.  The bound states act like an intermediary in the
process, which makes it also easier to understand that the properties
of the bound states play an important role in the production.  In the
non-perturbative regime, on the other hand, the electron-positron
pairs are created by the tunneling of the initially occupied Dirac sea
states into the quasibound states, which are not localized in space at
all.  The properties of the bound states before diving into the
negative continuum and becoming the quasibound state, however, still
influence the pair creation processes.  This can be viewed as the
non-Markovian feature \cite{schmidt:1998:quantum,schmidt:1999:non} of
the production.

This enhancement may be detected in the laboratory using the
Bethe-Heitler process
\cite{augustin:2014:nonperturbative,lotstedt2008laser}, interacting a
strong laser pulse with a highly charged ion or a nucleus.  Because of
the screen effect in a highly charged ion, a nucleus with similar
charge as an ion, based on our results, will produce more
electron-positron pairs when interacting with the same laser pulse.
On the other hand, pair creation here is triggered by bound states in
a binding potential well.  Whereas in a strong magnetic field, the
energy spectrum of the system will also be discretized
\cite{landau:1981:quantum}.  The creation processes under this field
configuration might be amplified by these Landau levels.  Likewise,
the spin of the created electrons and positrons might play a role
under magnetic field.  Because of this internal degree of freedom the
enhancement effect may appear in different manifestations, but much
more systematic studies to test these conjectures are necessary.  We
will report on these in future works.

\acknowledgments

We thank Drs. Q.~ Su, Y.~ J.~ Li, M.~ Jiang and N.~ S.~ Lin for
helpful discussion at the onset stage of this work.  This work is
supported by the Science Challenge Project (Grant No. TZ2016005), the
National Natural Science Foundation of China (Grant Nos. 11520101003,
11827807, and 11861121001), and the Strategic Priority Research
Program of the Chinese Academy of Sciences (Grant No. XDB16010200).
QZL thanks the Alexander von Humboldt Foundation for the support of
this project.

\bibliography{localized_effect}

\end{document}